# Subsidence and capillary effects in chalks
## Subsidence et effets capillaires dans les craies
## Setzungen und Kapillareffekte in Kreide

Pierre DELAGE
*Ecole Nationale des Ponts et Chaussées, Soil Mechanics Research Centre (CERMES), Paris, France.*

Christian SCHROEDER
*Université de Liège, Engineering geology laboratory (LGIH), Belgium.*

Yu Jun CUI
*Ecole Nationale des Ponts et Chaussées, Soil Mechanics Research Centre (CERMES), Paris, France.*

ABSTRACT : Based on the concepts of the mechanics of unsaturated soils where capillary phenomena arise between the wetting fluid (water) and the non-wetting one (air), the subsidence of chalks containing oil (non-wetting fluid) during water injection (wetting fluid) is analysed. It is shown that the collapse phenomenon of unsaturated soils under wetting provides a physical explanation and a satisfactory prediction of the order of magnitude of the subsidence of the chalk. The use of a well established constitutive model for unsaturated soils allows a description of the hydro-mechanical history of the chalk, from its deposition to the oil exploitation.

RESUME : En se basant sur les concepts de la mécanique des sols non saturés où les phénomènes capillaires entre le fluide mouillant (l'eau) et non mouillant (l'air) ont une importance prépondérante, on étudie la subsidence des craies réservoirs emplis d'huile (fluide non mouillant) lors de l'injection d'eau (fluide mouillant). On montre que le phénomène d'effondrement des sols non saturés lors du remouillage fournit une explication prhysique et permet une prévision raisonnable de l'ordre de grandeur de la subsidence de la craie. L'utilisation d'un modèle constitutif des sols non saturés bien établi permet d'autre part une description complète de l'histoire hydro-mécanique de la craie, depuis son dépôt jusqu'à l'exploitation.

ZUSAMMENFASSUNG : In wasserungesättingen Böden spielen Kappillareffekte zwischen dem benetzenden (Wasser) und nichtbenetzenden (Luft) Fluidum eine Hauptrolle. Auf dieser Grundlage wird die Setzung eines mit Öl (nichtbenetzenden Flössigkeit) gefüllen kreidigen Speichergesteins, wärend Wasserinjektionen (benetzenden Flüssigkeit) untersucht. Der Zusammenbruch eines ungesättigten Bodens während der Wiederbenetzung gibt eine physikalische Erklärung der Setzungen und ermöglicht eine ordentliche Prognose des Grössenordnung. Mit Hilfe eines gut festgelegen Modells des ungesättingen Bodens ist eine vollkommene Beschreibung der hydromechanischen Geschiehte der Kreide, seit Ablagerung bis zur Olgewinnung möglich.

## 1 INTRODUCTION

The North Sea oil fields are affected by a subsidence phenomenon which is known to derive directly from the compaction of the 200 m thick chalk reservoirs located under a 3000 m thick clay overburden. The compaction is mainly due to the decrease of the pore pressure within the chalk induced by the oil extraction. The rate of compaction as a function of the depletion can partly be explained by the visco-elasto-plastic features of the chalk. In an attempt to reduce the magnitude and rate of the subsidence, sea water has been injected in the reservoir, in order to maintain the level of the pore pressure. In spite of this, compaction is still running.

In order to better understand the mechanisms of compaction, the sensitivity of the chalk to the nature of the saturating fluid has been considered by various authors (see Monjoie, Schroeder *et al.* 1985). In this paper, the interactions between water, oil and chalk are considered in the light of recent developments of the mechanics of unsaturated soils. In unsaturated soils, capillary effects occur between water (the wetting fluid) and air (the non-wetting one). In chalk reservoirs, oil is the non wetting fluid. This approach has already been adopted in some works dealing with reservoir rocks. Brignoli *et al.* (1995) proposed a model based on Bishop's concept of effective stress for unsaturated soil. As confirmed by the authors, this concept is known to be useful in problems related to shear strength. However, its limitations regarding the prediction of volume change properties have been shown very early (Jennings & Burland 1962). Piau & Maury (1994) also studied the mechanical effects of capillarity during water injection in chalk reservoirs. They proposed a model which will be considered later in this paper.

The mechanical behaviour of unsaturated soils is strongly affected by the variations of the degree of saturation (of water) $S_r$, which is defined as the ratio between the volume of water and the total pore volume of the soil. It has also been shown that the suction $S$, defined as the difference between the air pressure $p_a$ and the water pressure $p_w$ ($S = p_a - p_w$) is a key state variable in the development of constitutive relations for unsaturated soils. The other state variable is the mean net stress, defined as ($p - p_a$), where $p$ is the mean total stress (Fredlund & Morgenstern 1976).

This paper shows that the mechanical behaviour of a chalk containing various fluids and the one of an unsaturated soil at various saturations of water are fairly similar. This is done by comparing experimental results obtained independently in two different laboratories. From this observation, it is showed that the so-called collapse phenomenon of unsaturated soils (Jennings & Knight 1957) can explain the subsidence due to water injection in reservoir rocks. More generally, it is shown that a constitutive model for unsaturated soils developed in Barcelona (Alonso, Gens & Josa 1990) provides a satisfactory framework for describing the complete hydro-mechanical history of the chalk, taking into account the mechanical consequences of the capillary effects occurring between water and oil within the chalk.

2 CHARACTERISTICS OF THE MATERIALS

*2.1 Chalk*

Chalk is made up of skeletal debris of unicellular alguae called coccoliths. The coccoliths are often desaggregated into grains, which are actually crystals of calcite. Calcite is the main mineral of chalk, although some chalks may contain up to 50% silica. In this work, a pure chalk with less than 1% silica was used. Observation using the scanning electron microscope (SEM) showed that the average size of the calcite grains ranged from 1 to 2 µm, and that the radius of the pores between the grains were comprised between 0.1 and 10 µm. The chalk is from the upper Cretaceous (Campanian) period. Samples come from a quarry located near Liège (Belgium).

The total porosity of chalk can be very high, up to 50% and more, which leads to a dry bulk density often less than 14 kN/m$^3$. In spite of this high porosity, and probably due to the arrangement of grains and pore space morphology, the permeability is rather low: it is included between 0.5 to 2 md, i.e. for water k = 0.5 to 2 10$^{-8}$ m/s.

*2.2 Unsaturated soil*

Unsaturated soils include natural soils in arid regions, and compacted soils used in any country for the construction of embankments and fills. In this work, a silt called "limon des plateaux" coming from the village of Jossigny (east of Paris) was used. Tests were performed on laboratory statically compacted samples. The geotechnical characteristics of the silt are given in Table 1. It is a low plasticity silt, located very close to the "A line" in the Casagrande classification. The clay fraction is 34%. Clay minerals, determined by X-ray diffractometry, are illite, kaolinite and interstratified illite-smectite.

SEM observation of a sample compacted at Proctor optimum ($w = 18\%$, $\gamma_d = 16,7$ kN/m$^3$) showed a rather dense arrangement of silt grains with an average diameter of 20 µm. Mercury intrusion pore size distribution measurements (Delage *et al.* 1996) showed that the more represented pore population had an entrance diameter close to 0.6 µm, with some poorly sorted pores with diameter comprised between 1.6 and 40 µm. As a consequence, the permeability to water of this soil is low, around 10$^{-9}$ m/s.

Table 1. Characteristics of the Jossigny silt.

| Liquid limit : $w_L = 37\%$ | Plastic limit : $w_P = 19\%$ |
|---|---|
| Plastic index : $I_p = 18$ | Porosity : $n = 39\%$ |
| % < 2 µm : 34% | % > 80 µm : 4 % |
| Proctor optimum water content : $w_{opt} = 18\%$ ||
| Proctor optimum dry unit weight : $\gamma_{d\ opt} = 16.7$ kN/m$^3$ ||
| Unit weight of the solid $\gamma_s = 27.2$ kN/m$^3$ ||

Both chalk and silt are made of grains. However, the grain size distribution of the chalk appeared to be much more well sorted (between 1 and 2 mm) than the one of the silica grains of the silt (between 4 and 50 µm). Furthermore, the silt contains 34% of clay particles, which are tightly coated to the grains at the Proctor optimum water content, and not very apparent in SEM.

# 3 VOLUME CHANGE BEHAVIOUR DURING ISOTROPIC COMPRESSION TESTS

## 3.1 Chalk

The influence of the saturating fluid on the volume change behaviour of the chalk submitted to an increasing isotropic stress is shown in Fig. 1, which presents the volume changes as a function of the logarithm of the stress for samples of chalk saturated with various fluids:
- 1. Water in chemical equilibrium with chalk
- 2. A mixture of water and sea water
- 3. Sea water
- 4. Kerosene (Soltrol)
- 5. Liquid paraffins of different viscosities (20 cp to 120 cp)

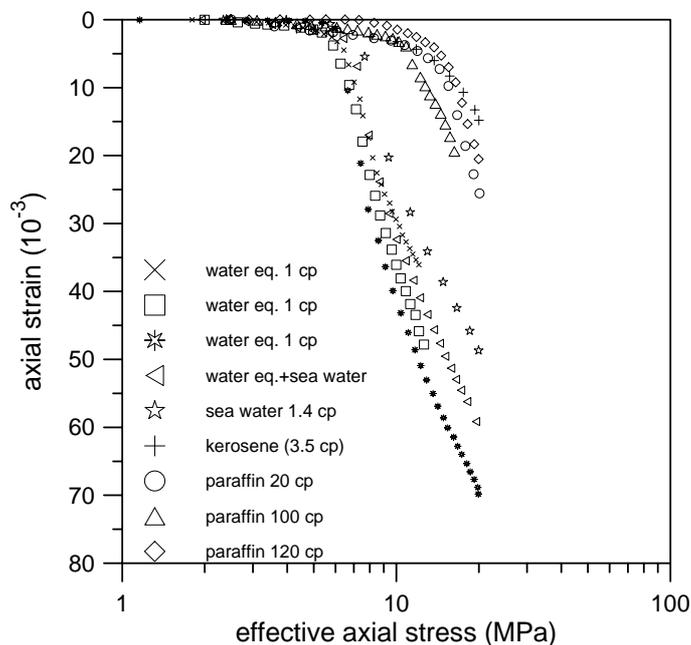

Fig. 1 : Isotropic compression tests on the samples of chalk saturated with various fluids.

Clearly, two groups of compression curves are observed :
- All the curves of the samples containing water are grouped together. Above 5-7 MPa, samples exhibit a drastic volume decrease induced by the collapse of the pores of the chalk (Schroeder 1995), evidencing quite a large compressibility.
- All the curves of the samples containing paraffin or kerosene, with viscosities ranging from 3.5 to 120 cp, exhibit a smaller compressibility than with water, they start compressing above 10-15 MPa.

Since no influence of viscosity is observed, the difference between the two behaviours should rather be related either to the wetting properties, or to the polarity of the pore fluid.

As in soils, two zones can be defined on the curves : a pseudo-elastic zone (also called over-consolidated zone) at lower stresses, where strains are reasonably reversible, and a plastic zone (corresponding to normally consolidated soils) at higher stresses. Fig. 1 shows that the elastic limit is displaced towards higher stresses when water is replaced by "oil", i.e. paraffin or kerosene.

## 3.2 Unsaturated soil

Fig. 2 presents the results of isotropic suction controlled compression tests made in the triaxial apparatus (between 0 and 0.6 MPa) on the Jossigny silt compacted at the Proctor optimum (Cui & Delage 1996). Tests were performed at various suctions ($S = 0,2 / 0,4 / 0,8$ and 1.5 MPa) corresponding respectively to degrees of saturation $S_r$ of 77, 73, 68 and 58%. $S_r$ decreases when suction increases.

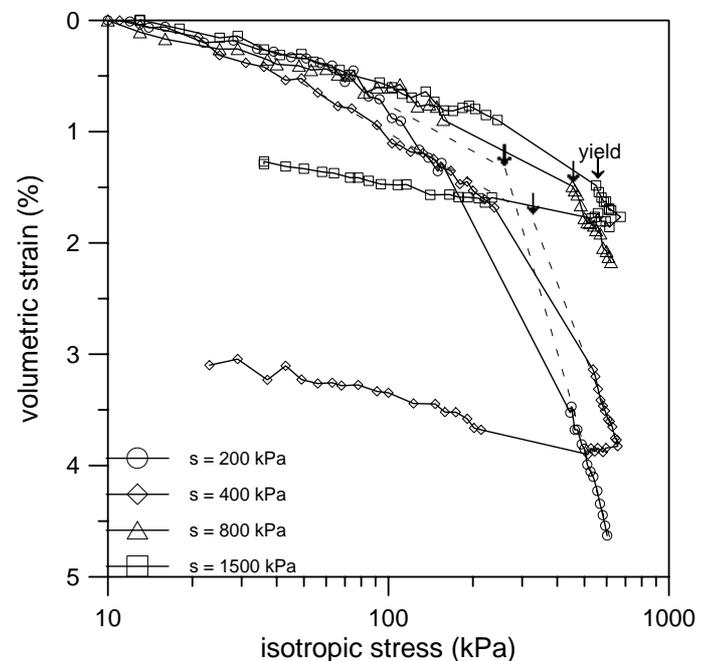

Fig. 2 : Isotropic constant suction compression tests on Jossigny silt samples, corresponding to various saturations of water.

Fig. 2 clearly shows that samples containing more water (suctions of 200 and 400 kPa) are more compressible than the dryer ones (suctions of 800 and 1500 kPa). Although the maximum applied stress (600 kPa) is somewhat low, one can separate the curves into two linear segments, corresponding to a pseudo-elastic behaviour at low stresses, and to a plastic behaviour at higher stresses. Irreversibility is confirmed by the unloadings at $S = 1500$ and 400 kPa.

Figs. 1 and 2 evidence a similar behaviour feature in chalk and unsaturated soil : an increase of the quantity of non-wetting pore fluid strengthens the material. The limit of the pseudo-elastic zone is displaced towards higher stresses, and the compressibility, defined as the slope of the curves in the volume-stress plane, decreases. This corresponds to a hardening phenomenon.

## 4 SHEAR BEHAVIOUR

### 4.1 Chalk

Axial stress-strain curves of standard triaxial tests under a constant confining stress (Montjoie *et al.* 1990) evidenced quite a high stiffness at small strains, and showed that failure was clearly defined by a peak or by an apparent change of slope. These features progressively disappear at higher confining stresses. This behaviour is typical of brittle materials, and has also been observed in sensitive clays (Graham *et al.* 1983). This is also consistent with the observations of Leroueil & Vaughan (1990), who showed that the mechanical behaviour of natural soils and weak rocks are quite similar. For low stresses, the initial microstructure of the material remains intact and gives reversible responses. For higher stresses, destructuration occurs and gives rise to plastic behaviour.

Considering that the peaks of axial stress-strain curves define the limit of the elastic behaviour, it is possible to determine an elastic zone in the mean effective / deviatoric stress plane ($p'/q$ with $q = \sigma_1 - \sigma_3$). This has been done in Fig. 3, for chalk samples filled with various pore fluids :
- 1. Water in chemical equilibrium with chalk.
- 2. Sea water
- 3. Dodecane (viscosity of 1.35 cp)
- 4. Oil (viscosity of 6.9 cp)
- 5. Liquid paraffin (viscosity of 120 cp)

Although some natural variability affects the legibility of the diagram, it is possible to draw various yield curves:
- The smaller elastic zone is observed on the sample containing sea water, which hence has the smaller resistance.
- The elastic zone of the sample containing water in a chemical equilibrium with chalk is slightly larger.
- Samples containing oil and dodecane have similar yield curves.
- The sample containing paraffin and the dry sample have similar yield curves and exhibit the stronger resistance.

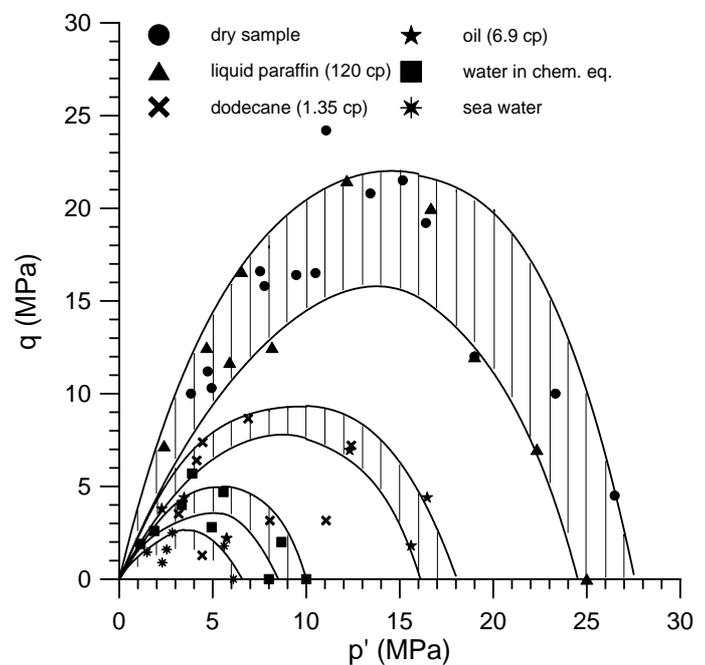

Fig. 3 : Effect of the pore fluid on the yield curves of the chalk in a q/p' diagram.

The yield curves of Fig. 3 have an elliptical shape with a small axis perpendicular to the p axis, and are fairly symmetrical. This means that the mechanical response of the chalk is fairly isotropic. This should be related to the relatively small effective stress supported during the stress history.

The enlargement of the elastic zone evidenced in Fig. 3 is consistent with the conclusions related to the influence of paraffin and kerosen drawn from the compression tests of Fig. 1. However, the reason why the yield curves of the samples full of dodecane and oil are in an intermediate position between the other curves is not obvious and deserves further investigation. This point is related to the nature of the chalk-fluid interaction at a microscopic level, which is obviously fluid dependent.

### 4.2 Unsaturated soil

Fig. 4 presents the yield curves of the unsaturated Jossigny silt at various controlled suctions, from 200 to 1500 kPa, represented in a $q / (p-p_a)$, where $(p-p_a)$ is the mean net stress. It shows, in a similar manner than Fig. 3, the enlargement of the elastic zone at larger suctions, for drier samples. This is a confirmation of the suction hardening phenomenon defined in the Barcelona model. Furthermore, the inclination of the yield curves shows the mechanical anisotropy of the compacted soil, induced during the $K_0$ compaction in the laboratory. Suction hardening is isotropic.

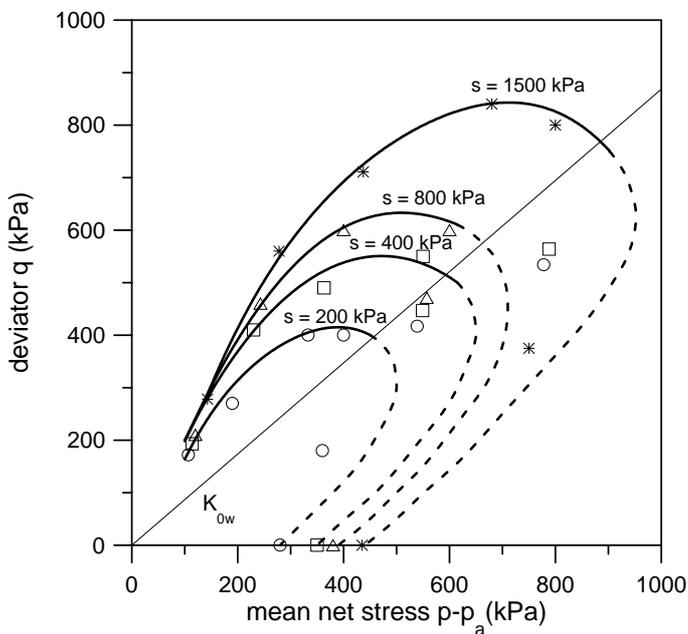

Fig. 4 : Effect of an increased suction on the elastic zone of the unsaturated soil in a *q/(p-p$_a$)* diagram.

## 5 THE COLLAPSE OF UNSATURATED SOILS ; APPLICATION TO CHALK RESERVOIRS

The previous similarities concerning the effects of the pore fluid on the mechanical properties of chalk and unsaturated soils can be summarised as follows: samples containing more water are weaker and more compressible, and liquids like paraffin or kerosene have on the chalk a strengthening effects similar to that of air in unsaturated soils.

### 5.1 The collapse of unsaturated soils

In unsaturated soils, the strengthening observed in drier soils is related to capillary effects, which are quantified by the value of the suction. Currently, unsaturated soils are schematically represented by the sketch of Fig. 5, where the soil particles are represented as roughly spherical grains, with air water menisci at contacts. This is somewhat simplistic in the case of fine grained soils, since the physico-chemical interactions between the molecules of clay and water are not considered. The air-water menisci exerts an increasing attraction between grains when suction is increased, resulting in a stronger macroscopic mechanical resistance of the arrangement of grains. This local attraction allows the stability of rather loose arrangements, which could not exist without menisci. When the soil is soaked, the menisci are destroyed, no more capillary attraction exists at the contacts, the loose structure of grains is destroyed, and the large pores initially existing between the grains collapse.

Collapse often occurs in loose natural unsaturated aeolian deposits like loess when wetted for first time, generally as a consequence of human activity (new constructions, leaks from canals or structures built on collapsible loess, very frequent in Eastern Europe, see Abelev & Abelev 1979). Due to their mode of deposition, aeolian soils are very recent and never supported any overburden. This is the reason why they are loose (porosities as high as 50%), slightly plastic ($I_p$ <10) and desaturate with little shrinkage, even in temperate countries in Europe. They are constituted of a loose arrangement of silt grains, and Fig. 5 provides a good approximation of their microstructure.

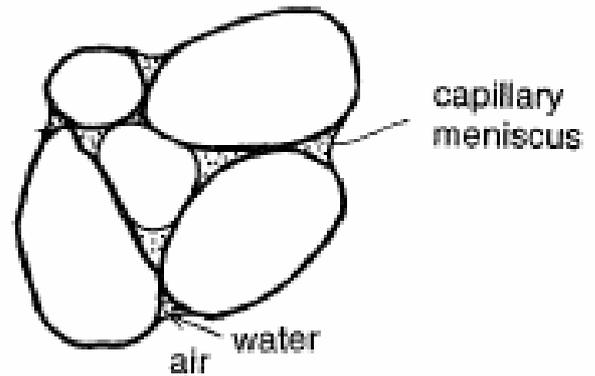

Fig. 5 : Schematical representation of unsaturated soils.

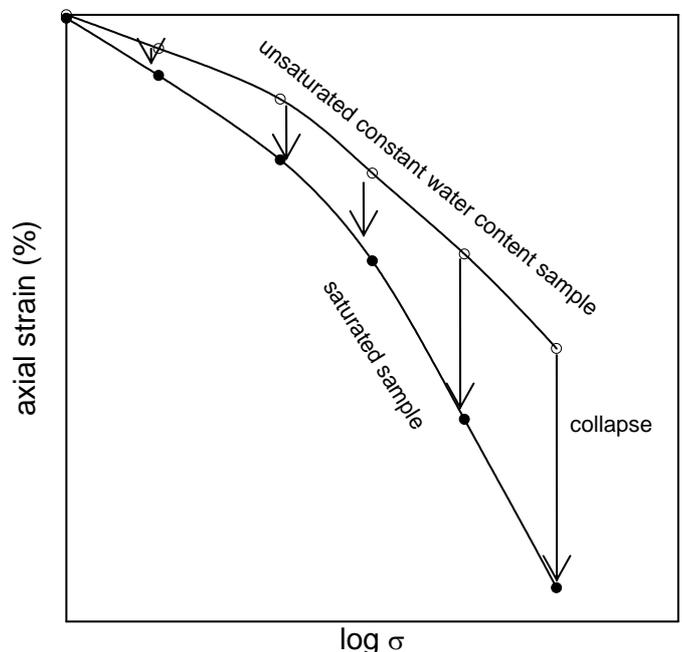

Fig. 6 : Testing the collapse potential of unsaturated soils.

Generally, the collapse potential of soils is studied in the oedometer by soaking samples under various constant stresses. As illustrated on Fig. 6 experiments show that all the final points after soaking are located near the compression curve of the satu-rated sample of the same initial porosity. This means that the saturated state is the more stable state of a soil submitted to a given state of stress, and that it is not possible to reach points located below the saturated compression curve.

This feature is the base of the method of the double oedometer (Jennings and Knight 1957), which is currently used for estimating the collapse potential of soils. The method consists of making a first test at a constant moisture content equal to the initial one, and a second one on the soaked sample. The collapse settlement at any stress is estimated from the vertical distance between the two curves (see Fig. 6).

*5.2 Application to chalk reservoirs*

The history of the formation of chalk reservoir rocks can be described as follows : the debris of coccoliths fell on the sea floor, getting slightly compacted under their own weight. Later, when the chalk was progressively covered by the subsequent impermeable clay overburden (3 000 m thick in some oilfields of the North Sea), the total stress was mainly supported by the pore pressure of the saturating sea water which cannot be expelled. Very few compaction occured, resulting in porosity values as high as 50%. During the diagenesis, some boundings between the grains of calcite appeared. In a further stage, water was progressively replaced by oil, without major change of the pore pressure. Due to capillarity, the complete replacement of water by oil has never been achieved, and chalk always keeps some residual water. Since the wetting properties of water on chalk are stronger than that of oil, water menisci are located in the smaller pores at the contacts between grains, as illustrated in Fig. 5 for unsaturated soils.

Based on this similarity and on the common behaviour features described in parts 3 and 4, it is proposed to interpret the settlement of reservoir chalk during water injection as a collapse phenomenon, with oil replacing air as the non-wetting fluid. Microscopically, it means that the oil-water menisci progressively disappear, and that the stability of the very loose arrangement of chalk grains cannot be ensured anymore by local capillary effects.

According to Jennings and Knight, the more stable state for chalk in Fig. 1 is defined by the curves obtained with samples saturated with water. The collapse settlement can be deduced, at a given stress, by the vertical distance separating the curves with oil (paraffin and kerosen) from the curves with water.

For some North Sea oilfield, under a mean effective stress estimated at 10 MPa, a strain of 3 % is obtained from Fig. 1, leading to a settlement of 6 m for a 200 m thick chalk deposit.

In spite of its simplicity, this analysis gives a value of settlement in the order of magnitude of the observed ones. It has been estimated from available experimental data obtained on a belgian chalk of a similar porosity, from tests which were not directly aimed at studying this problem. Complementary tests following more closely the stress-path supported by the chalk, run on samples coming from the oilfield could probably improve the prediction.

## 6. THE LC MODEL FOR UNSATURATED SOILS APPLICATION TO THE STRESS HISTORY OF A RESERVOIR.

*6.1 Introduction to the LC model (Alonso Gens & Josa 1990)*

Fig. 2 presented the isotropic compression curves of the compacted Jossigny silt at various controlled suctions. The curves were assimilated to two linear segments, the intersections of which being taken as yield points (Cui & Delage 1993). Results of Fig. 2 show that the yield stress is increasing with suction. This was interpreted as a suction hardening phenomenon, and confirmed by the yield curves of Fig. 4.

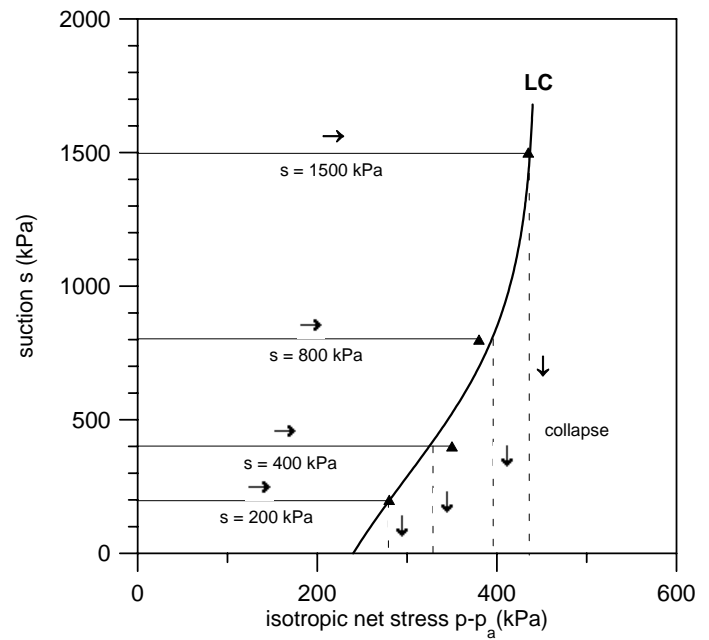

Fig. 7: The LC curve of the Jossigny silt in a $S/(p - p_a)$ plane.

Representing the yield stresses in a suction versus net mean stress plane (Fig. 7), one gets the shape of the elastic zone in the S/p plane. The yield curve is called LC curve (Loading-Collapse) in the Barcelona model. Every path reaching the LC yield curve gives rise to plastic strain. The constant suction compression tests of Fig. 2 correspond to horizontal lines. Tests where the suction is decreased under a constant stress correspond to vertical paths. The collapse tests of Fig. 6, where the suction was reduced directly to zero by soaking, are of this latter category. This is the manner the LC model predicts collapse settlements as plastic strains.

*6.2 Application to the reservoir history*

The framework proposed by the LC model makes possible a full description of the hydro-mechanical history of the reservoir, starting from the chalk deposition in sea water, and finishing with the depletion followed by water injection in the reservoir.

Since oil in chalk and air in unsaturated soils play a similar role of non wetting fluid, an extended notion of suction is proposed, equal to the difference between the oil pressure ($p - p_a$) and the water pressure $p_w$ ($S_o = p_o - p_w$). In a similar manner, the net mean stress is taken as $p - p_o$.

Let us consider the $S_o/(p - p_o)$ plane of Fig. 8.

- 1. Deposition of the chalk in sea water:

During this phase, there is a slight compaction of the chalk, followed by the set-up of diagenetic inter-grains bonds, corresponding to a hardening phenomenon. Since no oil was present in the chalk at that time, $S_o$ stays equal to zero and the stress path follows the net mean stress axis $p - p_O$. An initial elastic zone is defined on this axis between points O and A.

- 2. Infiltration of the oil within the chalk:

In this phase, one supposes that the penetration of oil occurs during the deposition of the clay overburden. Since oil penetrates the chalk, $p_o$ is bigger than $p_a$, and the suction $S_o$ is growing. If at the same time the deposition of the overburden occurs, there should be also a slight increase of the mean net stress. This path is represented by the AB segment in the figure. At that time, the chalk is hardening under the combined effects of increasing stress, diagenetic bonding and $S_o$ suction increase. In the LC model, the upper limit of the elastic zone is an horizontal line, called SI (Suction Increase) yield curve (see Fig. 8). The position of this line on the $S_O$ axis is defined by the highest suction supported by the sample. An upwards displacement of SI induces a displacement of LC towards the right. So, in agreement with experimental evidence in unsaturated soils, the AB stress path moves the LC yield curve towards the LC2 curve, on the right.

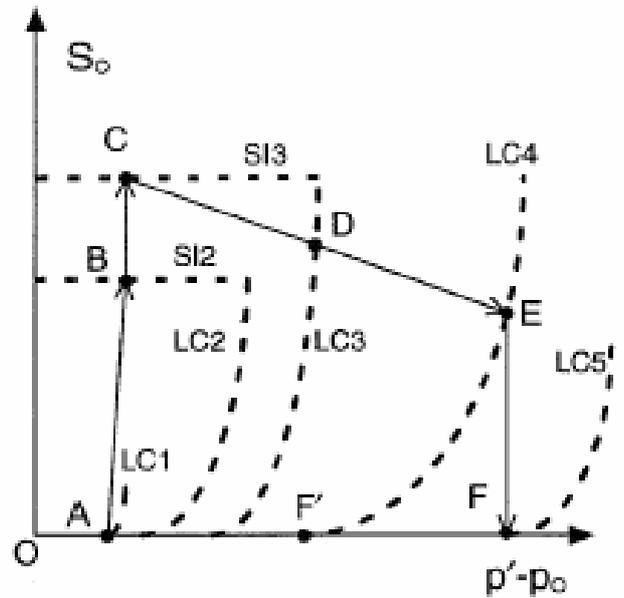

Fig. 8 : Hydro-mechanical history of the reservoir, taking into account the capillary effects.

- 3. Infiltration of oil within the chalk after the end of clay deposition:

The $S_o$ suction keeps on growing under a constant net mean stress. The stress path is now going vertically upwards. It moves SI towards SI3, and LC towards LC3. Point C corresponds to the end of the oil penetration in the chalk.

- 4. Depletion:

The reduction of the oil pressure $p_o$ under a constant total stress $p$ results in an increase of the net mean stress ($p - p_o$). The variation of $S_o = p_o - p_w$ during this phase also depends of the change of $p_w$, which cannot easily be predicted presently in the lack of relevant experimental data. However, results on unsaturated soils showed that suction decreases during compression tests at a constant water content. This is the reason why the corresponding stress path starting from C is going down towards the right. In a first stage (CD), elastic settlements occurs, followed by plastic ones when the LC curve is reached in D. The subsequent increase of mean net stress moves the LC curve towards LC4, up to the end of depletion, resulting in a hardening of the chalk due to porosity decrease.

- 5. Water injection:

During water injection, the change of the oil pressure $p_o$ and hence of the mean net stress ($p - p_o$) is not obvious. In an attempt of simplicity, it will be considered that $p_o$ after depletion is close to hydrostatic pressure, and that the water injection will not increase it significantly. This supposes a relatively high permeability, compatible with the high porosity of the chalk. Then, the mean net stress is likely to stay constant. Since the water pressure is increased, the $S_o$ suction decreases, and the stress path corresponding to water injection is a vertical line going downwards. This moves the LC yield curve towards LC5 (path EF), giving rise to the collapse plastic settlements estimated in part 5.2.

Water injection induces a hardening phenomenon by moving LC from LC4 to LC5, since the elastic zone is enlarged. It corresponds to a decrease of the void ratio, induced by the collapse settlement. In fact, according to the LC model and to experimental confirmations, the volume decrease occurring along the EF path is equal to the one which would occur along path F'F, since the initial and final positions of the yield curves are the same. In other words, the mechanical effect of the EF suction decrease is equivalent to the effect of a loading in the saturated state along F'F.

This observation provides an interpretation of the analysis of Piau & Maury (1994), who modelled the compaction of chalk during water flooding by using an additional local macroscopical stress called Π, acting at the water-oil front only. In fact, Fig. 8 shows that this approach can be included in a more general phenomenological understanding of the combined effects of capillarity and stress on the reservoir chalk. The local Π stress is in fact equal to the increment of net mean stress $(p-p_0)$ defined by the F'F segment.

## 7 CONCLUSION

By comparing experimental results coming from a reservoir chalk and an unsaturated soil, it has been shown that the mechanics of unsaturated soils provided a powerful framework for understanding the mechanical effects of water flooding on chalk reservoir rocks. Chalk compaction can be related to the collapse mechanism of unsaturated soils, and a technique used for the estimation of this collapse gives a satisfactory prediction of the settlements observed in some North Sea oilfields. A well established constitutive law for unsaturated soils also provides a complete description of history of the oilfield, including the mechanical effects of capillarity on volume changes.

ACKNOWLEDGEMENTS : Tests on chalk were performed within a joint research project supported by FNRS (Belgium), made by LGIH and the MSM Dept. (University of Liège). Exchanges between ENPC and ULg were made possible by the HCM European network ALERT. These organisations are gratefully acknowledged.